\documentclass[runningheads]{llncs}
\usepackage{graphicx}

\usepackage{booktabs}
\usepackage{amsmath,amssymb}
\begin{document}
\title{Volumetric Conditional Score-based Residual Diffusion Model for PET/MR Denoising}
\titlerunning{Volumetric Residual Diffusion Model for PET/MR Denoising}

%
%
\author{Siyeop Yoon \inst{1} 
\and Rui Hu \inst{1,2}  
\and Yuang Wang \inst{1,3} 
\and Matthew Tivnan\inst{1} 
\and Young-don Son\inst{4} 
\and Dufan Wu \inst{1} 
\and Xiang Li\inst{1} 
\and Kyungsang Kim \inst{1}  
\and Quanzheng Li \inst{1\dag}} 

\authorrunning{Siyeop Yoon  et al.}
%
\institute{Department of Radiology, Center for Advanced Medical Computing and Analysis, Massachusetts General Hospital and Harvard Medical School, Boston, MA, USA\and
Zhejiang University, Zhejiang, China\and
Tsinghua University, Beijing, China \and
Neuroscience Research Institute, Gachon University of Medicine and Science, Incheon, Korea 
}

\renewcommand{\thefootnote}{\fnsymbol{footnote}}
\footnotetext[4]{Corresponding authors.}

%
\maketitle              
\begin{abstract}
PET imaging is a powerful modality offering quantitative assessments of molecular and physiological processes. The necessity for PET denoising arises from the intrinsic high noise levels in PET imaging, which can significantly hinder the accurate interpretation and quantitative analysis of the scans. With advances in deep learning techniques, diffusion model-based PET denoising techniques have shown remarkable performance improvement. However, these models often face limitations when applied to volumetric data. Additionally, many existing diffusion models do not adequately consider the unique characteristics of PET imaging, such as its 3D volumetric nature, leading to the potential loss of anatomic consistency. Our Conditional Score-based Residual Diffusion (CSRD) model addresses these issues by incorporating a refined score function and 3D patch-wise training strategy, optimizing the model for efficient volumetric PET denoising. The CSRD model significantly lowers computational demands and expedites the denoising process. By effectively integrating volumetric data from PET and MRI scans, the CSRD model maintains spatial coherence and anatomical detail. Lastly, we demonstrate that the CSRD model achieves superior denoising performance in both qualitative and quantitative evaluations while maintaining image details and outperforms existing state-of-the-art methods. 

\end{abstract}
\section{Introduction}
Positron emission tomography (PET) has been widely employed in various realms due to its ability to provide insights into molecular-level activities, contingent upon the design of radioactive tracers and regional uptake. However, the signal-to-noise ratio (SNR) of PET is inherently limited by various factors, including the restricted dosage of the radiotracer and the patient radiation exposure. Consequently, this limited SNR may compromise both the diagnostic quality and the speed of PET imaging. One practical approach to enhance the SNR of PET images is through post-processing techniques aimed at reducing noise while preserving the underlying signal. Among the various post-processing methods, filtering-based techniques such as Gaussian and total variation (TV) denoising are commonly used but often struggle to capture the full characteristics of PET images, leading to significant blurring in the results. 
\begin{figure}[t]
\centering
\includegraphics[width=0.7\textwidth]{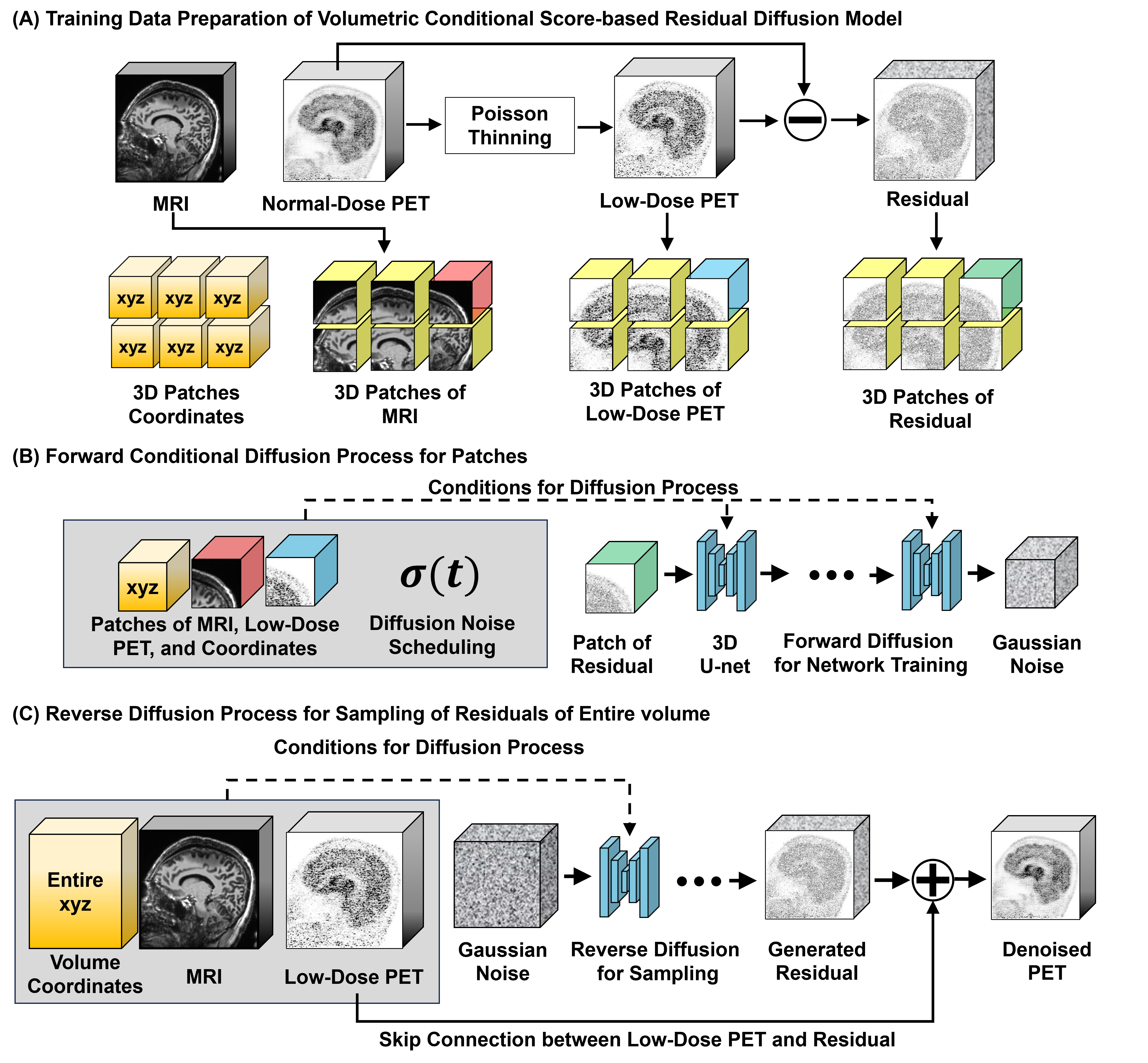}
\caption{\textbf{A overview of volumetric PET denoising using a conditional score-based residual diffusion model}. (A) The normal-dose PET volumes undergo Poisson thinning to simulate a low-dose PET scan scenario. Subsequently, the residual of normal-dose and low-dose PET volumes is generated. The residual, low-dose PET, and MRI volumes were split into smaller patches along with their respective spatial coordinates. (B) During the forward diffusion process, these patches undergo a noise addition process with time-dependent scheduling, represented by $\sigma(t)$. Then, the 3D U-net is trained for a score-matching function by removing an additive Gaussian noise with patch conditions of low-dose PET, MRI, and coordinates. (C) The trained network samples the residual of the entire volume from the Gaussian noise via the reverse diffusion process conditioned by entire low-dose PET and MRI associated with coordinates.} \label{fig1}
\end{figure}

Recent advancements in deep learning-based methods have demonstrated superior performance in PET image denoising. These methods leverage the non-linearity of deep learning models and their ability to incorporate anatomical information from other modalities such as MRI or CT, enhancing the denoising performance. The most widely utilized deep learning architectures for supervised denoising models are based on U-shaped convolutional neural networks (CNNs) \cite{gong2018pet,schaefferkoetter2020convolutional,liu2022personalized} and transformers\cite{jang2022pet,jang2023spach}. However, these supervised models face challenges, such as the requirement for high-quality reference images and mismatches in data distribution between the training and testing phases, which can degrade the overall quality of the denoised images. Alternatively, deep image prior techniques leverage PET and anatomical pairs through deep neural networks without necessitating high-quality PET images \cite{kim2018penalized,gong2018pet2,gong2021direct}. However, anatomical structures and spatially inhomogeneous noise distribution can lead to spatially varying denoising performance. Furthermore, the aforementioned deep learning methods are deterministic, which limits the uncertainty analysis of denoising performance.

The diffusion model, specifically for PET imaging denoising \cite{gong2023pet,shen2023pet,shen2024bidirectional}, offers notable outcomes and enables uncertainty analysis through diffusion process realizations. This is because the diffusion model learns the data set's log density gradient, called the score function, and provides the stochastic representation of the dataset \cite{ho2020denoising,song2020score,EDM}. Despite its promising performance and utilities, the computational burden limits its adoption in practice. The diffusion model is inherently slow and requires huge computation resources due to the iterative denoising process during inference. Furthermore, PET imaging is essentially three-dimensional data; it requires a computationally efficient model capable of representing volumetric data while existing diffusion models easily exceed the memory capacity, which causes an additional level of difficulty for model development. 

In this study, we present an efficient 3D conditional score-based residual diffusion (CSRD) model designed explicitly for enhancing PET image denoising. Our model utilizes the EDM framework \cite{EDM,wang2023patch}, optimized for 3D data handling and efficient computational performance. By refining the score function representation and applying 3D patch-wise training, our approach effectively addresses the challenges of the computing cost and memory demands in the 3D diffusion model. Our methodology not only facilitates rapid volumetric denoising of 3D PET images within three minutes but also significantly boosts denoising performance with an MR prior.

\section{Method}
\subsection{Score-based Diffusion model for Residuals}
In this section, we adapt the score-based diffusion model framework to focus on the residual distribution between pairs of images to generate residuals to denoise the image instead of generating the image (Fig. \ref{fig1}). Our objective is to establish a mapping from Gaussian noise to the residual distribution, which may not necessarily follow Gaussian or Poisson distributions.

Let us denote the residual between low-dose PET volume $x_\text{Low}$ and normal-dose PET image $x_\text{Nor}$ as $r=x_\text{Low} - x_\text{Nor}$. Initially, consider a family of mollified distributions $p(r;\sigma, x_\text{Low})$ obtained by adding i.i.d. Gaussian noise of standard deviation $\sigma$ to the residuals. For a maximum noise level $\sigma_{max}$, $p(r;\sigma_{max}, x_\text{Low})$ is nearly indistinguishable from pure Gaussian noise. The diffusion process can be described using a probability flow ordinary differential equation (ODE), where the evolution of a residual $r_a \sim p(r_a;\sigma(t_a))$ from time $t_a$ to $t_b$ results in a residual $r_b\sim p(r_b;\sigma(t_b))$. 

The ODE, as adapted for the residual distribution, is given by \cite{EDM}: 
\begin{equation}
    dr = -\dot{\sigma}(t)\sigma(t) \nabla_r \log p(r; \sigma(t)) dt 
\end{equation}
Here, $\nabla_r \log p(r; \sigma(t))$ represents the score function for the distribution of residuals, pointing towards higher density regions at a given noise level $\sigma(t)$. The score matching for residuals can be expressed as:

\begin{equation}
   L(D_\theta;\sigma) = \mathbb{E}_{y \sim p_{res}} \mathbb{E}_{n \sim \mathcal{N}(0, \sigma^2 I)} [D(y + n; \sigma) - y]^2 
\end{equation}
This defines a loss function for the score-based diffusion model, where $y$ represents the residual sampled from the distribution of residuals $p_{res}$, and $n \sim N(0,\sigma^2I)$ is the noise added to the residual. The denoising score-matching function $D$ is a neural network parameterized by $\theta$ that aims to predict the residual from the noisy version of it. The loss measures the mean squared error between the estimated residual and the true residual as:
\begin{equation}
    \nabla_r \log p(r; \sigma, x_\text{Low}) = \frac{D(r; \sigma) - r}{\sigma^2}
\end{equation}
The loss function guides the score-matching neural network in training to estimate the residuals accurately at different steps of the reverse diffusion process. By adapting these equations to focus on the residual distribution, the diffusion model framework can effectively model the transition between image pairs through the distribution of their residuals, providing an indirect tool for image-to-image translation tasks.

\begin{figure}[t]
\centering
\includegraphics[width=.9\textwidth]{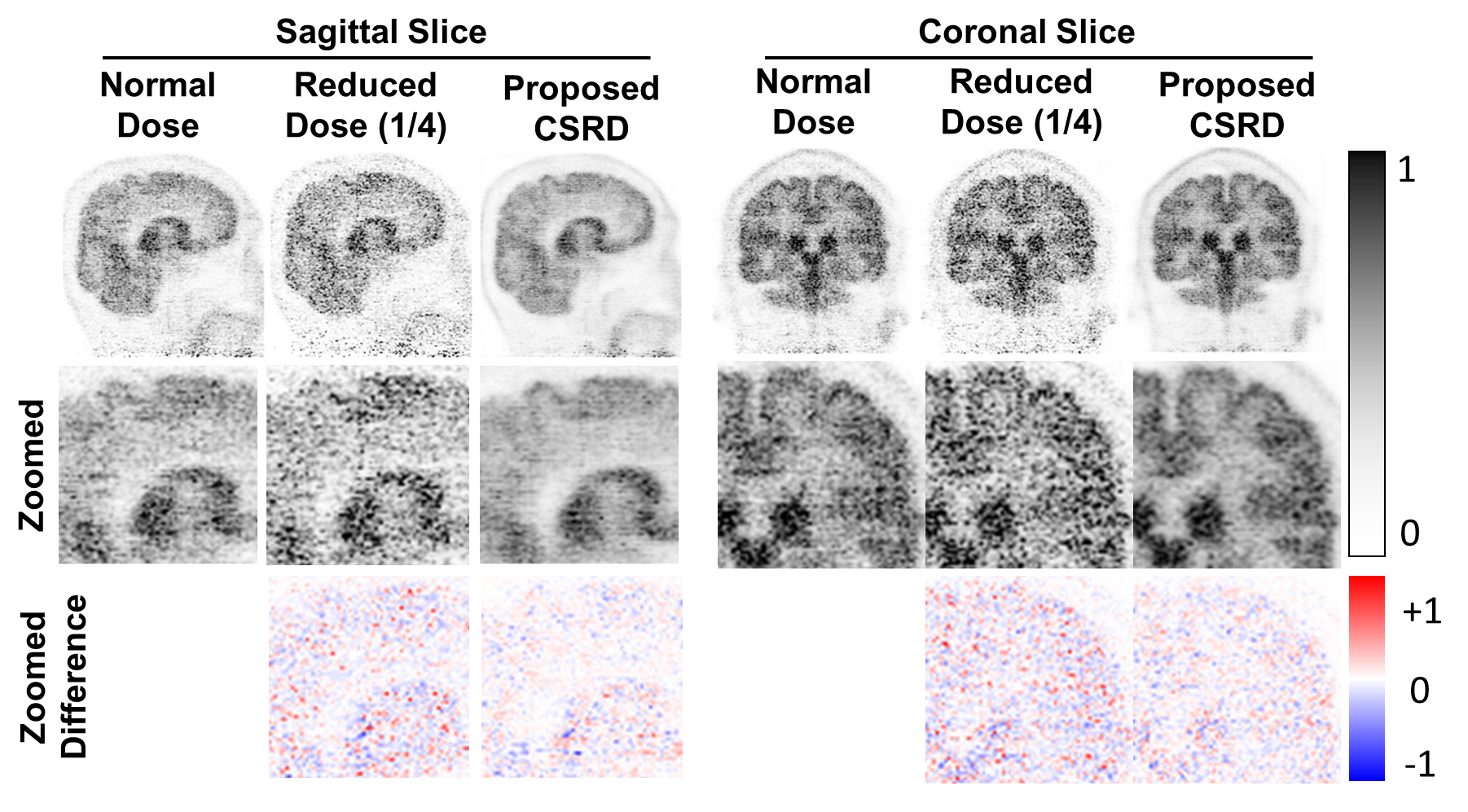}
\caption{\textbf{The representative PET images}. The proposed conditional score-based diffusion model successfully improved the quality of PET volume while preserving fine anatomical details.} \label{fig2}
\end{figure}
\subsection{Patch-wise Training of Volumetric Conditional Score-based Diffusion Model}
We will integrate a prior distribution into the residual-based diffusion process to extend the adaptation of the diffusion model framework for PET image denoising, conditioned by low-dose PET and MR images. By incorporating structural priors of the MR image, we aim to enforce known anatomic constraints during the diffusion process.

Let us now consider a prior distribution $p(x_\text{MR})$ representing knowledge about the anatomy. During the diffusion processes, the model aims to reduce the noise (estimate accurate residual $r$) and ensure that the image adheres to the prior distribution. The modified objective can be expressed as follows:
\begin{equation}
\label{ODE}
    dr =  -\dot{\sigma}(t){\sigma(t)} \nabla_r \log p \left( r ; \sigma(t),  x_\text{Low}, x_\text{MR}\right) dt
\end{equation}

Here, $\nabla_r \log p \left( r; \sigma, x_\text{Low}, x_\text{MR} \right)$ is the score function that now also takes into account the prior knowledge about anatomy by MR images. This gradient guides the reverse diffusion process towards solutions that are more consistent with the prior anatomy distribution. This can be achieved by conditioning the neural network on low-dose PET and MRI images. During training, the score-matching function learns to balance noise reduction with adherence to the prior, optimizing the overall anatomical coherence in the denoised image.

Incorporating MRI priors into the 3D residual diffusion process ensures that the denoised images maintain anatomical integrity and consistency. However, training a 3D diffusion model is resource-intensive due to the significant memory requirements and the necessity for large 3D datasets. To mitigate these challenges, a 3D patch-wise loss approach is employed, allowing the model to learn from smaller, uniformly sampled sub-volumes. This approach capitalizes on the shift-invariance property of convolution layers, allowing the model to generalize well across different spatial locations in the data, thus making the training process more feasible and efficient without compromising the learning quality.

To adapt the 3D patch-wise loss function for a conditional score-based residual diffusion model, consider that the model is conditioned not only on the noise level but also on a prior that describes the 3D location of patches. Then, the 3D patch-wise loss is given as

\begin{equation}
    L_P(D_\theta;\sigma) = \sum_{i=1}^N \mathbb{E}_{y \sim p_{res}} \mathbb{E}_{n \sim \mathcal{N}(0, \sigma^2 I)} \| D(y_{\Omega_i}+n; \sigma,x_\text{Low}, x_\text{MR}, \Omega_i )-y_{\Omega_i}\|^2_2
\end{equation}
, where  $\Omega =\bigcup_{i=1}^N \Omega_i$ and $N$ is a total number of sub-divisions of 3D coordinate of low-dose PET volume $\Omega$. Once the model is trained, the trained model is used to estimate the residuals of low-dose PET images using the reverse-ODE flow (Eq. \ref{ODE}) with the second-order corrected stochastic reverse diffusion sampling \cite{EDM}, which enables further reduction in the reverse diffusion process.

\section{Experiments}
\subsection{Dataset Description}
The brain PET dataset scanned 27 individuals with a 7T MRI and HRRT-PET scanner (Siemens Medical Solutions, Knoxville, TN)\cite{kim2015serotonin}. The study protocol was approved by the Institutional Review Board of the local institute and all procedures used in the study were conducted in accordance with international ethical standards as set out in the Declaration of Helsinki. Each participant received an injection of 11 C-DASB tracers, with an average dose of 577.6 $\pm$ 41.0 MBq. These scans were performed to quantify cerebral serotonin transporter binding. The PET images, with a matrix size of 256 × 256 × 207 and voxel size of 1.21875 × 1.21875 × 1.21875 mm$^3$, were reconstructed using the 3D-OP-OSEM algorithm. The corresponding T1-weighted MR images were collected with specific parameters: repetition time of 4000ms, echo time of 5.26ms, flip angle of 10 degrees, inversion time of 900ms, and a voxel size of 0.7 × 0.7 × 1.5 mm$^3$. The MR images were resampled to the identical spatial resolution as the PET images; then, PET images were cropped in the same field of view of MRI, resulting in a matrix size of 160 × 160 × 160. 

We divided 27 patients randomly into a training group (20 patients) and a testing group (7 patients). The training and testing were conducted using normal-dose PET volume as a reference and low-dose images, reconstructed using a 3D-OP-OSEM algorithm from reduced counting via Poisson thinning \cite{kim2018penalized} as the model input. We applied the Poisson thinning for downsampling directly to the raw data (like listmode) without any adjustments, such as normalization, random, scatter, and attenuation corrections. The assumption is that the scatter and random fractions remain unchanged in low-count data. The Poisson thinning was specifically used to downsample regular dose data, allowing for random discarding of coincidence events based on a set sampling factor. For training, 3D PET volumes were reconstructed at 4×, 6×, and 8× reduced doses in a total of 60 pairs of low-dose and normal-dose volumes together with T1-weighted MR images. Also, testing low-dose 3D PET volumes were generated at 4×, 6×, 8×, and 10x downsampling levels. Notably, the 10× downsampled low-dose PET represents an unseen level of dose reduction during the model training.

\subsection{Implementation Details}

We implemented our diffusion model within a framework of the original implementation of EDM \cite{EDM}. The network architecture was modified for 3D patch-wise training by extending the network's layers into 3D. During the training, the patch size was 64$^3$ and randomly sampled. The model has inputted the patches of residuals, low-dose PET, T1-weighted MR, and coordinates. The number of channels was set to 64, and mixed precision was used due to computational efficiency. The model was trained using Adam optimizer with a learning rate of 0.002 and a minibatch size 16 in 4× NVIDIA A100 40GB GPUS (5 days) for 65k iterations. If not mentioned otherwise, the same hyperparameters as those in the original implementation were used. 

For performance evaluation of the proposed CSRD method, other most commonly used denoising techniques were also implemented, including Total variation denoising (TV), U-net\cite{ronneberger2015u}, and Restormer \cite{jang2022pet}. U-net and Restormer are U-shaped networks based on CNN and transformer, respectively. Both models had four levels of encoder-decoder, and the number of channels was 64. Both models were implemented in 2D and inputted a slice of low-dose PET and T1-weighted MR, which have a matrix size of 160 × 160. The learning rate was set to 0.002, and the minibatch size was 16. The training of U-net and Restormer was performed with MAE loss using Adam optimizer in one NVIDIA A100 40GB GPUS for 65k iterations. In addition, we separately trained our CSRD model without a T1-weighted MR image to investigate the impact of anatomic prior. Our source code and trained models will be available online. 

We evaluated the performance of CSRD with multiple metrics, including mean absolute error (MAE), peak-signal-to-noise ratio (PSNR), Structure similarity index (SSIM), Haralick feature distance ($H_{dist}$) \cite{wu2017iterative}, and perceptual distance \cite{zhang2018unreasonable} ($P_{dist}$). The Haralic feature distance measures the similarity of texture as follows: 

\begin{equation}
    H_{dist}(x_\text{Nor},x_\text{Denoised}) = \sqrt {\sum_{i=1}^N (\frac{h_i (x_\text{Denoised}) -h_i (x_\text{Nor})}{h_i (x_\text{Nor})}}) ^2,
\end{equation}
where $h_i (x_\text{Nor})$ and $h_i (x_\text{Denoised})$ are i-th components of the Haralick texture feature of the normal-dose and denoised low-dose images, respectively. The perceptual distance is defined by the mean-squared error of deep features as:
\begin{equation}
    P_{dist}(x_\text{Nor},x_\text{Denoised}) = \frac{1}{N}\sum_{i=1}^N (VGG_i (x_\text{Denoised}) -VGG_i (x_\text{Nor}))^2,
\end{equation}
where $VGG (\cdot)$ is a deep features extracted through a pretrained VGG-19 model. 


\section{Results}
\begin{figure}[t]
\centering
\includegraphics[width=\textwidth]{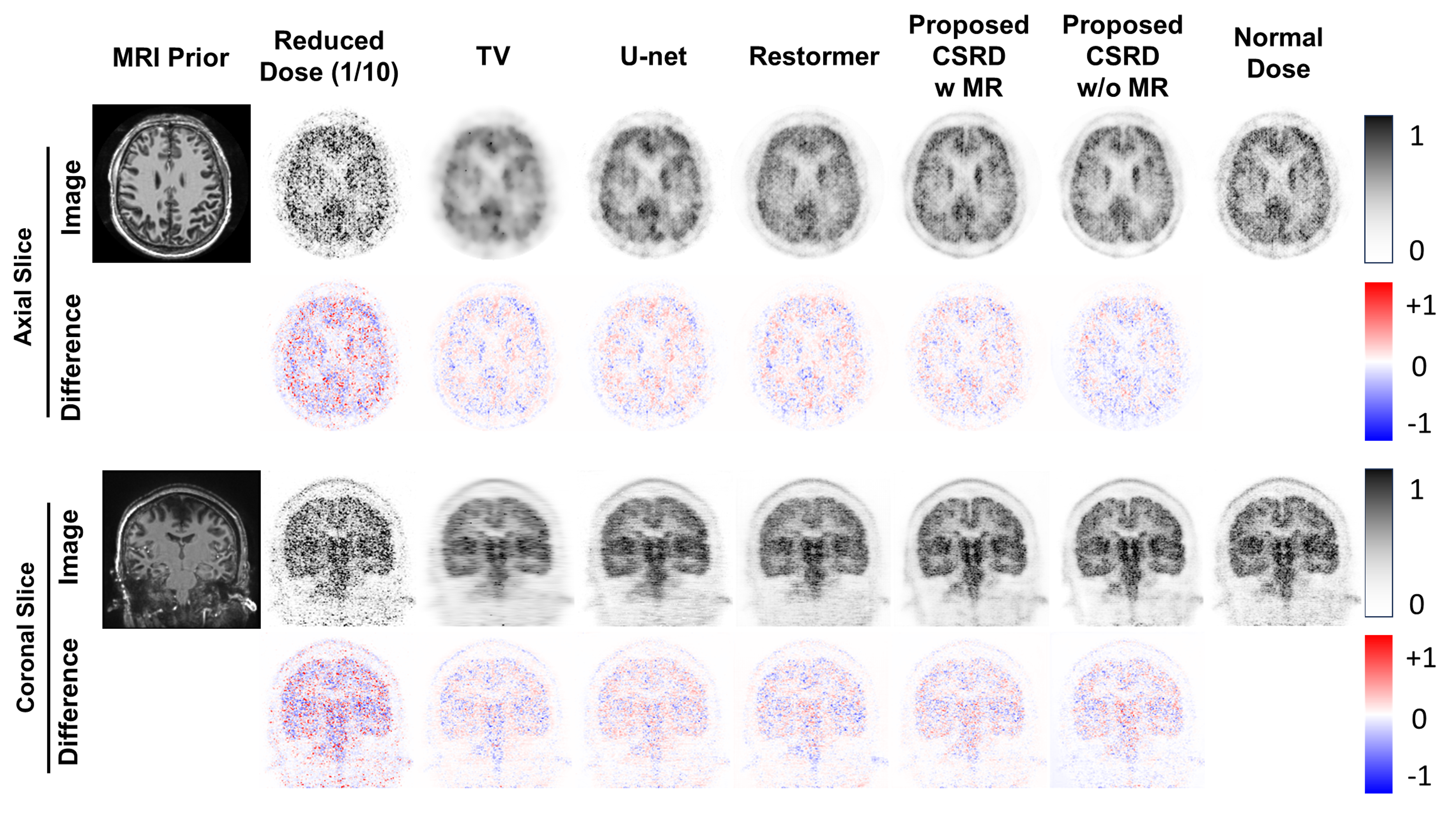}
\caption{\textbf{The images and error map in the unseen level of noise (1/10-dose)}} \label{fig3}
\end{figure}

We showcase the performance of the proposed CSRD model for 3D PET denoising in Fig \ref{fig2}. We set the number of function evaluations to 100 for inference, and our model took 3 minutes to denoise 3D PET with a volume size of 160 × 160 × 160 and required only 12GB on a single GPU. Table \ref{tab} summarizes the quantitative results of denoising performance compared to original low-dose and other SOTA denoising techniques. The presented data suggest that deep learning-based methods outperformed TV filter-based denoising techniques due to their ability to capture more complex textures. Additionally, our diffusion model outperforms all other models in terms of performance in both traditional image metrics and feature-based metrics. Figure \ref{fig3} demonstrates the denoised images and error map in the unseen level of noise (1/10-dose). Even if the dose was not included in the training, our proposed model minimized blurring through axial slices and showed anatomical consistency because it leveraged 3D information. In the case of CSRD without MRI, similar performance was shown compared to the results with MRI, but residual noise components in the background that were not removed can be confirmed. This suggests that the anatomical prior in our model provides additional performance improvement by estimating the residuals with a conditional diffusion process.

\begin{table}[htbp]
  \caption{Quantitative Comparison of the 11 C-DASB PET Denoising Performance }
\makebox[\textwidth][c]{
    \begin{tabular}{ccccccccccccc}
\toprule
Methods     & MAE ($\downarrow$)                         & PSNR ($\uparrow$)               &SSIM ($\uparrow$)           & $H_{dist}$($\downarrow$) &&&$P_{dist}$ ($\downarrow$) \\ \midrule
Low-Dose    & 0.059 $\pm$ 0.026             & 35.07$\pm$ 4.21    & 0.84 $\pm$0.070  & 4.84 $\pm$ 1.74 &&& 0.27 $\pm$ 0.17   \\
TV          & 0.036 $\pm$ 0.015             & 40.04$\pm$ 4.39    & 0.91 $\pm$0.048  & 22.53 $\pm$ 8.59 &&& 0.33 $\pm$ 0.22\\
U-net       & 0.034 $\pm$ 0.014             & 40.08$\pm$ 4.36    & 0.92 $\pm$0.042  & 4.57 $\pm$ 2.39 &&& 0.18 $\pm$ 0.12\\
Restormer   & 0.036 $\pm$ 0.015             & 40.65$\pm$ 4.68   & 0.92 $\pm$ 0.045  & 3.59 $\pm$  1.31 &&& 0.14 $\pm$ 0.06 \\\hline
\textbf{CSRD w MR}        & \textbf{0.033 $\pm$ 0.015 }   & \textbf{41.11$\pm$ 4.97}    & \textbf{0.93 $\pm$0.044 }  & \textbf{2.78 $\pm$ 1.21} &&& \textbf{0.10 $\pm$ 0.06 } \\ 
CSRD w/o MR & 0.056 $\pm$ 0.014             & 40.56$\pm$ 4.56    & 0.91 $\pm$ 0.039                     & 3.21 $\pm$ 1.68 &&& 0.10 $\pm$ 0.06\\

\bottomrule
\end{tabular}%
}
\label{tab}%
\end{table}%

\section{Conclusion and Discussion}

In this study, we proposed the conditional score-based residual diffusion model (CSRD) for efficient volumetric PET denoising. Through a refined score function representation and 3D patch-wise training, the model significantly reduces the computational load and accelerates the denoising process. By integrating volumetric data from PET and MRI scans, the CSRD model preserves spatial coherence and anatomical details, enhancing the PET image's diagnostic quality. The CSRD model demonstrated superior denoising performance compared to other state-of-the-art methods, as evidenced by quantitative evaluations across various metrics. Further studies are warranted in the assessment of the proposed method in other radiotracers, such as [$^{18}$F]FDG and [$^{18}$F]MK-6240, as well as different modalities for anatomic priors in PET/CT. We assumed that the registration of MR and PET can be achieved using clinical PET-MR scanners. However, additional research is needed to investigate changes in denoising performance resulting from image registration. Notably, PET images typically do not exhibit Gaussian noise, yet in 3D patches with uniform intensities, residual distribution more closely resembles Gaussian distribution, highlighting the potential application of CSRD in other image modalities.

\bibliographystyle{splncs04}
\bibliography{reference}

\end{document}